\documentclass[twocolumn,prl,showpacs,preprintnumbers,amsmath,amssymb]{revtex4}

\usepackage{graphicx}
\usepackage{dcolumn}
\usepackage{bm}
\usepackage{subfigure}

\begin{document}

\preprint{}

\title{Observation and absolute frequency measurements of the $^{1}S_{0}$-$^{3}P_{0}$ optical clock transition in ytterbium}

\author{C. W. Hoyt}
\email{hoycha@boulder.nist.gov}
\author{Z. W. Barber}
\altaffiliation{University of Colorado, Boulder, CO, 80309}
\author{C. W. Oates}
\author{T. M. Fortier}
\altaffiliation{Los Alamos National Laboratory, Physics Division, MS
H803, Los Alamos, NM, 87545}
\author{S. A. Diddams}
\author{L. Hollberg}
\affiliation{National Institute of Standards and Technology\\325
Broadway, Boulder, CO 80305}
\thanks{Official contribution of the National Institute of Standards and Technology; not subject to copyright.}

\date{\today}

\begin{abstract}
We report the direct excitation of the highly forbidden $(6s^{2})
^{1}S_{0} \leftrightarrow (6s6p) ^{3}P_{0}$ optical transition in
two odd isotopes of ytterbium. As the excitation laser frequency is
scanned, absorption is detected by monitoring the depletion from an
atomic cloud at $\sim$70 $\mu$K in a magneto-optical trap. The
measured frequency in $^{171}$Yb (F=1/2) is 518,295,836,593.2 $\pm$
4.4 kHz. The measured frequency in $^{173}$Yb (F=5/2) is
518,294,576,850.0 $\pm$ 4.4 kHz. Measurements are made with a
femtosecond-laser frequency comb calibrated by the NIST cesium
fountain clock and represent nearly a million-fold reduction in
uncertainty. The natural linewidth of these J=0 to J=0 transitions
is calculated to be $\sim$10 mHz, making them well-suited to support
a new generation of optical atomic clocks based on confinement in an
optical lattice.
\end{abstract}

\pacs{32.30.Jc, 06.30.Ft, 32.80.Pj, 39.30.+w}

\maketitle


Work is underway to realize a high-performance optical clock that
combines the best features of state-of-the-art single-ion
and neutral atom 
optical frequency standards \cite{Diddams04}. Such a clock system
comprises a dipole-force optical lattice trap that confines an
ensemble of neutral atoms individually to sub-wavelength sites for
Doppler- and recoil-free precision spectroscopy
\cite{Katori03,Porsev04}. Crucial to this scheme is an atom with a
``clock'' transition that has both a narrow linewidth and an
insensitivity to lattice perturbations. Several groups have
recognized ytterbium as an excellent candidate and are pursuing
lattice-based optical clocks that will use its narrow $^{1}S_{0}$
$\leftrightarrow$ $^{3}P_{0}$ transition
\cite{Park03,Maruyama03,Porsev04,Hong05}. However, until now the
absolute frequency of this resonance was known in tables
\cite{Martin78} to only a few gigahertz. We report the direct
excitation of the doubly-forbidden $(6s^{2}) ^{1}S_{0}$
$\leftrightarrow$ $(6s6p) ^{3}P_{0}$ optical transition at 578.4 nm
in two odd isotopes of ytterbium \cite{Cramer05}. Using a
femtosecond-laser frequency comb \cite{Udem99}, we make precision
absolute frequency measurements with an uncertainty of 4.4 kHz -- an
improvement of nearly 10$^{6}$. This accurate frequency knowledge
will expedite the pursuit of an ytterbium-based optical clock, whose
performance can potentially surpass that of the best cesium primary
standard \cite{Jefferts04} by orders of magnitude.

\begin{figure}[htb]
\centerline{\includegraphics[width=8.3cm]{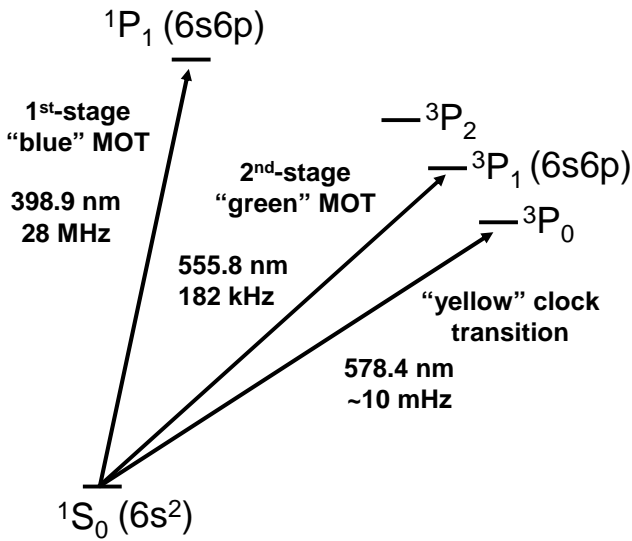}}
 \caption{Ytterbium atomic energy levels. Wavelengths and natural linewidths are indicated for the
 relevant cooling, trapping, and clock transitions. The fine-structure splitting is not to scale.
 The hyperfine structure for the $^{171}$Yb (I=1/2) and
 $^{173}$Yb (I=5/2) isotopes is ignored for clarity.}
 \label{levels}
\end{figure}

The ytterbium $^{1}S_{0}$ $\leftrightarrow$ $^{3}P_{0}$ resonance is
an outstanding potential clock transition, in part because of its
narrow natural linewidth ($\sim$10 mHz \cite{Porsev04}). This
transition is strictly dipole-forbidden from spin and orbital
angular momentum considerations. An appreciable excitation
probability exists, however, through hyperfine mixing of the
$^{3}P_{0}$ level with nearby states in the odd isotopes. (See the
diagram in Fig.\ \ref{levels}.)

In an optical atomic clock based on confinement to a lattice, the
lattice laser wavelength is chosen such that the ac Stark shift of
the upper clock energy state matches the shift of the ground state.
This results in a vanishing net perturbation to the clock frequency.
The shift-canceling wavelength for ytterbium is calculated to be 752
nm \cite{Porsev04}, readily accessible by high-power cw
titanium-sapphire laser systems. Furthermore, the J=0 to J=0
transitions in Yb reported here are expected to depend minimally on
lattice polarization \cite{Porsev04}, enabling straightforward
implementation of two- and three-dimensional lattice geometries.
However, experimental measurements of the shift-canceling lattice
wavelength are required to rule out possible deleterious effects of
higher-order processes.

A lattice-based neutral atomic frequency reference has several
important features. In addition to a high signal-to-noise ratio (due
to averaging signals from $\gtrsim$10$^{5}$ atoms) the lattice
enables long spectroscopic interaction times (1 s) and accompanying
narrow-linewidth measurements. These two factors lead to high short-term 
stability. Also, the high accuracy of an ion system may be reached
since its motional state control can be mimicked for each neutral
atom in its respective lattice site. With each atom tightly confined
to a fraction of the lattice wavelength (i.e., in the Lamb-Dicke
regime), systematic frequency uncertainties associated with the
Doppler shift are essentially eliminated \cite{Bergquist87}.
Additionally, if a three-dimensional lattice has an occupation
number less than one, collisional frequency shifts are expected to
be negligible \cite{Katori03}.

Offering both high stability and high accuracy, this
best-of-both-worlds frequency reference scheme was first proposed
for $^{87}$Sr \cite{Katori02}. Recently, its analogous clock
transition has been observed and measured \cite{Courtillot03}, and
high-resolution Doppler-free spectroscopy of lattice-confined
$^{87}$Sr atoms has been demonstrated \cite{Takamoto03}. One of the
primary differences between Sr and Yb is their hyperfine structure.
The nuclear spins of $^{87}$Sr, $^{171}$Yb and $^{173}$Yb are I=9/2,
1/2, and 5/2, respectively. The larger total angular momentum of
$^{87}$Sr leads to lower temperatures \cite{Xu03,Maruyama03}, but
may introduce unwanted experimental issues such as optical pumping
and increased sensitivity to lattice polarization for the clock
transition \cite{Katori03}. Ytterbium offers two odd isotopes with
reasonable natural abundance ($^{171}$Yb: 14\%, $^{173}$Yb: 16\%)
and the possibility to explore the relatively simple I=1/2 spin
system. Ytterbium has been laser-cooled and trapped
\cite{Loftus00,Park03,Maruyama03} and researchers have achieved
Bose-Einstein condensation of $^{174}$Yb by all-optical means
\cite{Takasu03b}.

As a first step toward a lattice-based system, we investigated the
ultranarrow clock transition in a cloud of cold atoms. Ytterbium can
be cooled to low temperatures because its energy level structure
(see Fig.\ \ref{levels}) is amenable to two successive
stages of laser cooling and trapping as described below. 
Lower temperatures aid in the search for absorption on the
ultranarrow clock transition because smaller Doppler-widths lead to
higher-contrast signals. Courtillot, \textit{et al.}\
\cite{Courtillot03} used a novel approach to find the $(5s^{2})
^{1}S_{0} \leftrightarrow (5s5p) ^{3}P_{0}$ clock transition in
$^{87}$Sr by monitoring the trap fluorescence from 2 mK atoms in a
first-stage magneto-optical trap (MOT). 
The clock transition was observed as a 1 \% depletion in
fluorescence when the clock laser frequency was on resonance. Here
we cool an Yb atomic sample to tens of microkelvins with the
addition of a second-stage MOT using the 555.8 nm transition shown
in Fig.\ \ref{levels}. Then with 20 mW of power at 578.4 nm focused
to a spot slightly larger than the size of the atomic cloud
(1/e$^{2}$ beam radius of $\sim$1 mm), we drive the clock transition
at an estimated Rabi rate of $\sim2\pi$ $\times$ 4 kHz. A single
$\pi$-pulse addresses approximately 2 \% of the atoms in a 70 $\mu$K
sample ($\Delta\nu_{Doppler}\approx240$ kHz,
full-width half-maximum). 
Adding successive $\pi$-pulses during the lifetime of the trap, we
are able to address more atoms and increase trap depletion to over
80 \%.

As shown in Fig.\ \ref{levels}, a blue transition at 398.9 nm offers
a broad transition (natural linewidth: $\Gamma = 2\pi$ $\times$ 28
MHz) for a first-stage MOT. We obtain temperatures of $\sim$3 mK for
$^{171}$Yb and $^{173}$Yb, and $\sim$7 mK for $^{174}$Yb. Similar to
sub-Doppler cooling in the alkaline-earths \cite{Xu03}, the lower
temperatures of the odd isotopes can be explained by the presence of
hyperfine ground-state structure \cite{Maruyama03}. The cooling
laser source is a system of InGaN laser diodes
\cite{Park03,Komori03}. Two slave diodes are injection-locked by a
master external-cavity diode laser. The frequency of the light from
each slave laser is controlled
independently through the use of acousto-optic modulators. 
The master laser is locked to the desired isotope by saturation
spectroscopy using an Yb hollow-cathode lamp \cite{Kim03}. The
output from one of the slave lasers is red-detuned from resonance by
$\delta\equiv\omega_{laser}-\omega_{0}\approx-4\Gamma$ and acts as a
slowing beam as it counter-propagates with respect to the atomic
beam. 
A second slave laser produces $\sigma^{+}-\sigma^{-}$ trapping beams
(red-detuned to $\delta=\Gamma/1.75=-2\pi$ $\times$ 16 MHz) for the
MOT that uses a magnetic field gradient of $\sim$4.5 mT/cm. All of
the beams are fiber-coupled to the vacuum chamber region to provide
spatial filtering. The slowing beam power is 4 mW, and the powers of
the horizontal and vertical trapping beams are 2 mW and 1 mW,
respectively. Under these conditions, up to $\sim$2 $\times$
10$^{6}$ atoms can be trapped in the blue MOT.

Atoms from the blue MOT are transferred to a 555.8-nm second-stage
MOT with up to 65 \% efficiency. As shown in Fig.\ \ref{levels}, the
natural linewidth of this $^{1}S_{0} \leftrightarrow ^{3}P_{1}$
transition is $\Gamma=2\pi\times182$ kHz, which implies a
$\sim$150-fold decrease in the Doppler cooling limit with respect to
the 398.9 nm transition. Measured temperatures are then as low as 30
$\mu K$ for $^{173}$Yb, with trap lifetimes as long as 3 s. The
second-stage cooling laser light is provided by a narrow linewidth
fiber laser system (1 W at 1111.6 nm) that is frequency-doubled.
Second-harmonic light at 555.8 nm with a power of 30 mW is generated
in a single pass through a 5 cm MgO-doped periodically-poled lithium
niobate crystal \cite{Pavel04,Wallerand04}. This light is then split
into trapping, probe and spectroscopy (locking) beams. After
fiber-coupling to the chamber area, up to 4.5 mW is measured in each
trapping beam with $1/e^{2}$ radii of $\sim$4 mm. The frequency is
red de-tuned several linewidths from resonance. The
frequency-modulated spectroscopy beam intersects the atomic beam at
normal incidence downstream from
the trapping area. 
Phase-sensitive detection of the green fluorescence is used to
stabilize the IR master oscillator to the desired isotope. 

The yellow excitation light at 578.4 nm is provided by a highly
stable dye laser. The laser is spectrally narrowed using a tunable
high-finesse reference cavity, which is in turn locked to an
ultrastable cavity with a finesse of greater than 150,000
\cite{Young99}. Some of the frequency-correction capabilities were
removed from this system to maximize power for spectroscopy,
resulting in a laser linewidth of $\sim$5 kHz. The longitudinal
modes of the ultrastable cavity drift at a rate of a few hundred
hertz/day and therefore serve as an effective reference in the
search for the extremely narrow $^{1}S_{0}$ $\leftrightarrow$
$^{3}P_{0}$ clock transition. Ultimately, we found the resonance in
both $^{171}$Yb and $^{173}$Yb by applying the 578.4-nm clock
radiation while the atoms were in the green MOT. However, the
absorption lineshape is Stark-shifted by $\sim$300 kHz due to the
presence of the green trapping beams.

To avoid this large Stark shift, we used a method based on a series
of excitation pulses that follows the demonstration and suggestions
contained in Ref.\ \cite{Courtillot03}. The blue MOT is loaded for
100 ms in the presence of the green MOT beams. The blue beams are
then turned off and the green MOT (dB/dz $\approx$ 1.5 mT/cm) is
allowed to settle for 30 ms. After turning off the green MOT beams
the fluorescence from a 200 $\mu$s resonant green pulse is detected
with a photomultiplier tube and integrated to serve as a
normalization factor for each measurement. A sequence of 300 $\mu$s,
578.4 nm yellow pulses (traveling or standing wave) is applied in a
nearly vertical direction while the trapping beams are off. The
pulses are separated in time by 1 ms, during which time the green
MOT is on. This allows the remaining trapped atoms to re-thermalize
(i.e., refill velocity holes) and the excited atoms to
gravitationally accelerate downward and out of resonance with the
next pulse. The green MOT magnetic field gradient is on during the
excitation pulses, although its effect on the J=0 transition is
negligible for this measurement ($\sim$10 kHz/mT \cite{Porsev04}).
After the prescribed number of yellow pulses (typically 50), the
integrated fluorescence from a final resonant green pulse serves to
read the atom depletion relative to the first normalization green
pulse.

\begin{figure}[htb]
\subfigure{\centerline{\includegraphics[width=8.3cm]{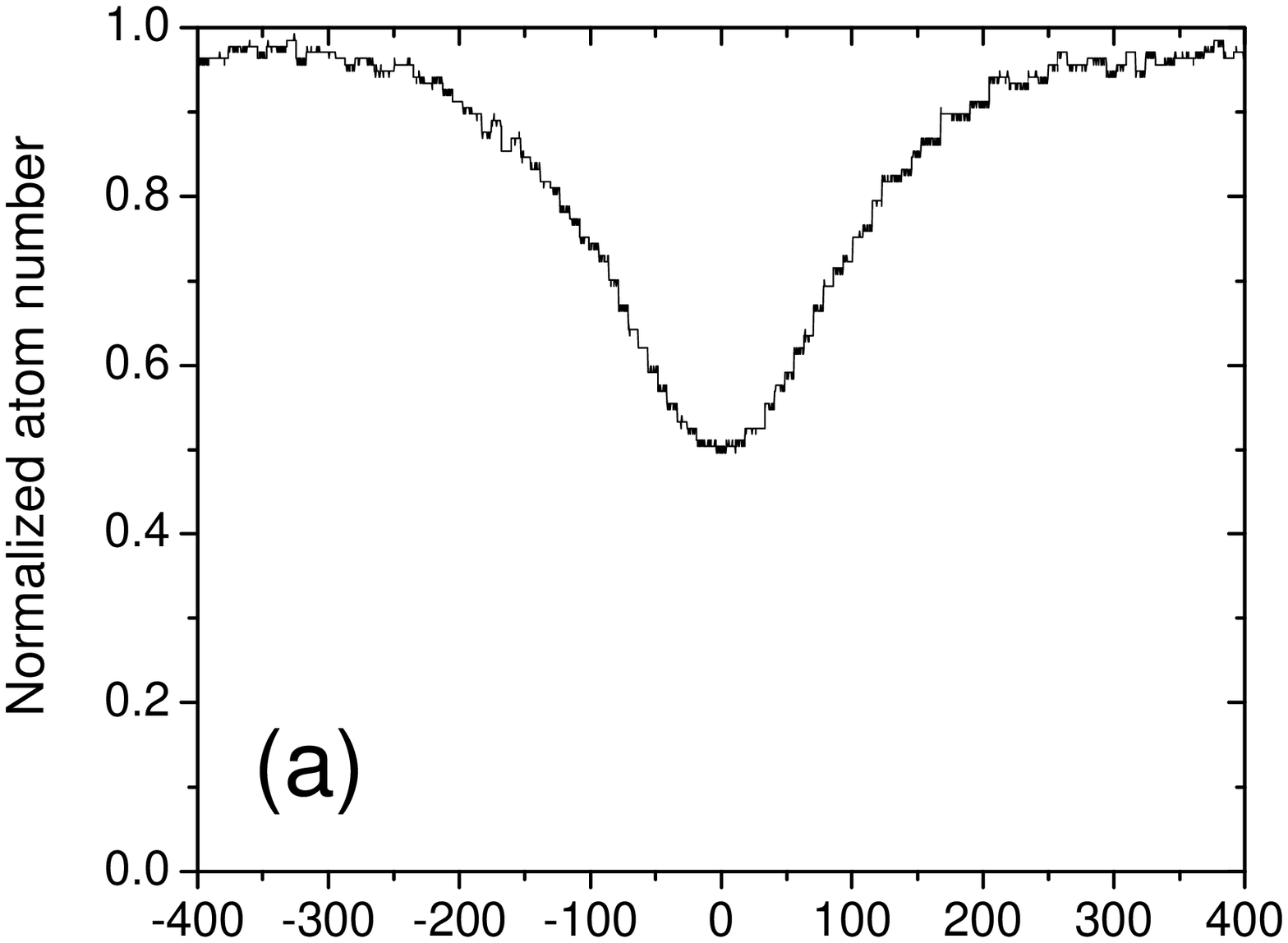}}\label{171data}}
\subfigure{\centerline{\includegraphics[width=8.3cm]{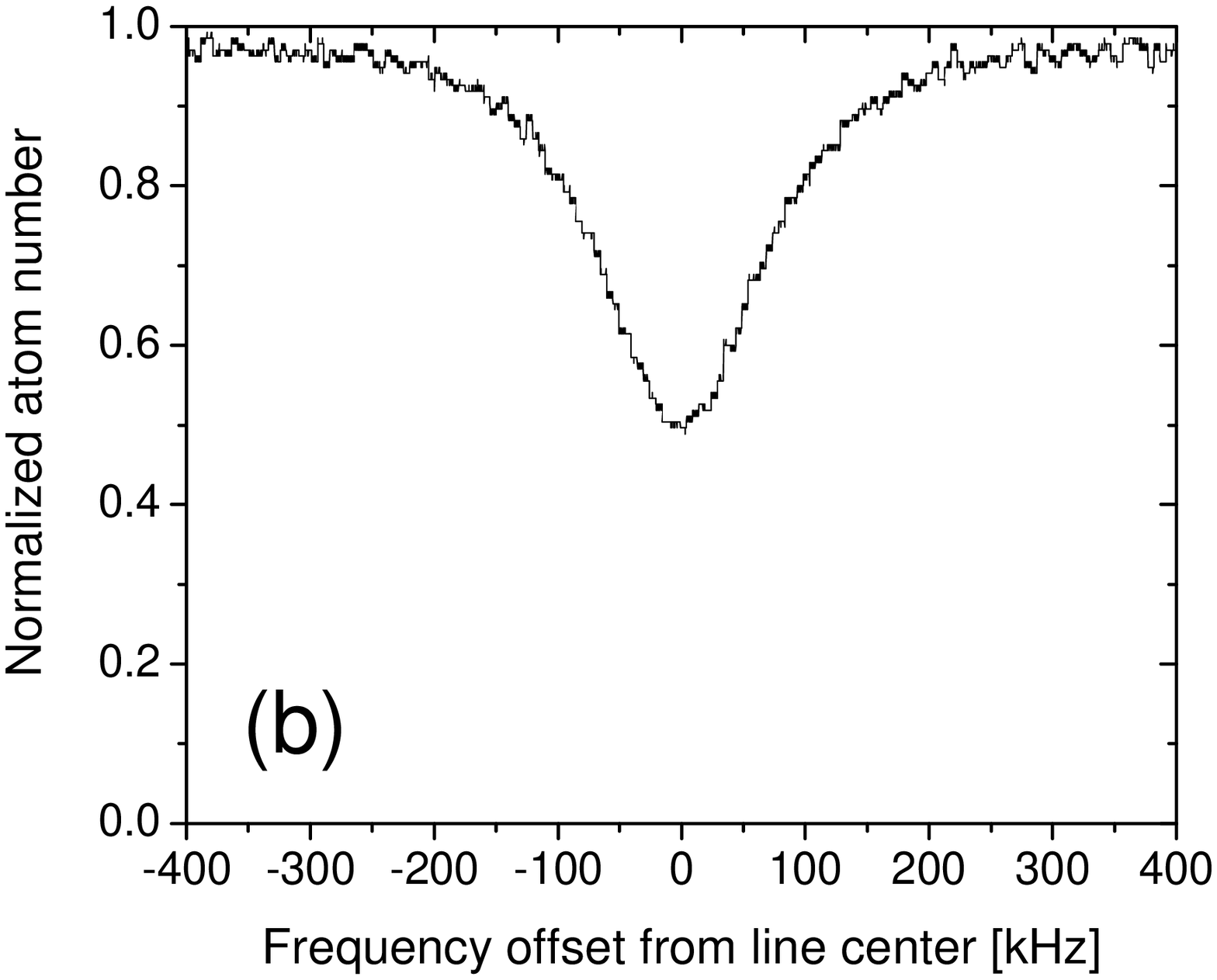}}\label{173data}}
\caption{Lineshape data for the $^{171}$Yb $^{1}S_{0}$
$\leftrightarrow$ $^{3}P_{0}$ (a) and $^{173}$Yb $^{1}S_{0}$
$\leftrightarrow$ $^{3}P_{0}$ (b) transitions, averaged four times
and taken with 50 traveling-wave pulses at each 7.5 kHz step. Each
step lasts $\sim$200 ms.}
\end{figure}


Figures \ref{171data} and \ref{173data} show excitation data for the
doubly-forbidden $^{1}S_{0}$ $\leftrightarrow$ $^{3}P_{0}$
transition in $^{171}$Yb and $^{173}$Yb, respectively. The ordinate
in the figures indicates the fraction of
trapped atoms relative to the off-resonance case. 
Lineshapes are determined primarily by the velocity distributions of
the atoms in the second-stage MOT, and are fit well by Gaussians.
The corresponding temperatures, 84 $\mu$K for $^{171}$Yb and 48
$\mu$K for $^{173}$Yb, agree well with time-of-flight measurements.
The difference in temperatures most likely results from the
difference in angular momenta for the two isotopes
\cite{Maruyama03,Xu03}.

The absolute frequencies of these resonances were determined from a
beat frequency between the stabilized cw excitation laser and
specific modes of an optical frequency comb. The comb was generated
from a broadband Ti:sapphire femtosecond laser oscillator, which
enables the measurement of the comb offset frequency without the use
of highly nonlinear fiber \cite{Bartels02,Ramond02}. Both the mode
spacing and the offset frequency of the self-referenced comb are
stabilized to a hydrogen maser that is calibrated by the NIST cesium
primary frequency standard. The mode-spacing of the comb is
determined by the 998731460 Hz repetition rate of the laser. The
measured frequency of the $(6s^{2}) ^{1}S_{0}$ $\leftrightarrow$
$(6s6p) ^{3}P_{0}$ transition in $^{171}$Yb (F=1/2) is
518,295,836,593.2 $\pm$ 4.4 kHz. This number represents the mean of
six measurements that have a standard deviation of 1072 Hz. The
measured frequency of the $(6s^{2}) ^{1}S_{0}$ $\leftrightarrow$
$(6s6p) ^{3}P_{0}$ transition in $^{173}$Yb (F=5/2) is
518,294,576,850.0 $\pm$ 4.4 kHz. This number represents the mean of
three measurements that have a standard deviation of 216 Hz. To
within the stated uncertainty, the isotope shift between $^{171}$Yb
and $^{173}$Yb is 1,259,743.2 kHz. This is comparable to the
corresponding isotope shift of 1,271,600 kHz measured for the
$^{1}S_{0}$ $\leftrightarrow$ $^{3}P_{1}$ transition at 555.8 nm
\cite{Clark79}.

Two tests allow us to limit Doppler-related uncertainties associated
with a possible atomic cloud drift velocity to 4.0 kHz. First, there
is no measurable shift in line centers between traveling- and
standing-wave configurations to within the fitting uncertainty.
Second, a saturation dip present in higher-resolution, standing-wave
lineshapes constrains drift-velocity shifts by indicating the
relationship between Doppler line-center and the laser frequency
associated with the zero-velocity atoms. We also note the presence
of another saturation feature that is likely due to the
near-vertical angle of the yellow spectroscopy laser beam. The
subtle effects of this geometry complicate accuracy evaluation:
during a single probe pulse there is non-negligible atomic
gravitational acceleration leading to effective chirping of the
laser frequency as seen by the atoms. Further investigation will be
required to fully understand the implications of this geometry.

Constrained to 4.0 kHz, the Doppler-related error dominates other
systematic factors. Including uncertainties associated with
lineshape fitting, ac Stark shifts from residual trapping laser
light, and comb measurements, the total measurement uncertainty is
4.4 kHz. This accurate knowledge of the absolute clock transition
frequencies in ytterbium will expedite Doppler-free spectroscopy in
a lattice -- an important step in the development of a new
high-performance optical atomic clock.

The authors wish to thank J.\ C.\ Bergquist for invaluable
assistance with the clock laser at 578.4 nm. They also appreciate
early experimental help by O.\ Ovchinnikov and femtosecond laser
development by A. Bartels. C.\ W.\ H.\ is grateful for support from
the National Research Council.


\end{document}